# EMPOWERED CUSTOMERS IN E-HEALTH BUSINESS PROCESS

M. Anshari[#1]   M.N. Almunawar[#2]

[#1, 2] *Faculty of Business, Economic & Policy Studies*
*Universiti Brunei Darussalam, Brunei Darussalam*

**Abstract-** E-health innovations support empowered customers. It offers the ability for customers to have greater control and ready access applications of health information, clinical information, and social interaction between interested groups. However, providing empowerment in any state of interaction levels to customers (patients) in a healthcare organization is challenging tasks. Customers are empowered in the sense of controlling the process of interaction between a firm with its customers, and among customers themselves. This paper discusses dimension of customers' empowerment in e-health business process. We propose reference model of Personal Health Cycle (PHC) as a holistic view of healthcare business process. The PHC is used to define and distinct *electronic health record* (EHR) from *electronic medical record* (EMR) and customers empowerment.

## INTRODUCTION

Empowerment in a healthcare organization characterizes by the manner in which all parties share responsibilities in the healthcare process that can be expected to pose challenges. The challenges for the healthcare providers were mainly (1) the capability of customers to interact with their systems online (B2C = "business to consumer"); (2) improved possibilities for institution-to-institution transmissions of data (B2B = "business to business"); (3) new possibilities for peer-to-peer communication of consumers (C2C = "consumer to consumer") (Eysenbach, 2001).

Empowerment through technological approaches such as e-health system can be used to provide effective self service also helps organization to reduce costs by handling an increasing number of customer transactions effectively. The outcome of empowerment relate to disease and treatment such as self management of disease and treatment, perceived control over the disease, satisfaction regarding the treatment, self determination of health and treatment, level of health literacy, etc. While, healthcare provider-patient interaction may include patients' satisfaction regarding consultation or intervention, number of self efficacy in asking question, shared decision making, level of preference regarding participation in consultation, etc.

The purpose of this paper is to propose the explicit scope of empowerment in e-health business process. In the next section, we present a literature review on related work and Section 3 contains the methodology of our research. We present discussion in Section 4, and finally Section 5 details the conclusions.

**EMPOWERMENT**

The idea of empowerment emerges because many discussions among researchers indicate that customers' empowerment in health services plays a critical role in achieving health outcome for themselves. In 1990s, a healthcare organization started to use the Internet based systems to promote self-management and educate consumers about health, wellness, healthcare options, and disease management strategies (Morris et al., 1997; Simmons, 2001). Gibson (1991) proposed that empowerment of individual is a social process of recognizing, promoting, and enhancing peoples' abilities to meet their own needs, solve their own problems and mobilize the necessary recourses in order to feel in control of their lives. While, Brennan and Safran, (2005) proposed empowerment is a characteristic of groups and individuals that energizes them with the knowledge and confidence to act in their own behalf in a manner that best meets identified goals. Public health empowerment can best be seen through the actions of self-help groups and collectiveness who claim the right to define health concerns in terms of those most affected by them rather than those who seek to care for them. Empowerment of individuals provides a useful starting point for examining how Consumer Health Informatics (CHI) enables consumers to actively manage their own health concerns and participate in their own care. Web led to the creation of health related Internet resources, to ensure direct access by consumers to professional and research biomedical literature and to commercial health information management providers such as WebMD.

Development in Web-enabled access to clinical records systems provided and opportunity for healthcare systems to provide patients with access to their clinical records, thus expanding the portfolio of CHI tools In that sense, empowerment characterizes the manner in which patients and clinicians approach care, with mutual expectations, rights, and responsibilities to assume a greater level of deliberate self-involvement in the care process. CHI innovations have the capacity to support empowerment of lay persons in managing their own health concerns and acquiring the necessary healthcare resources to achieve health goals. IT requirements to support empowerment include 4 key functions: access to comprehensive, reliable, and relevant health information; communication with peers and professionals; access to personal care management tools including self-monitoring and DSS; and ubiquitous access to clinical records (Brennan and Safran, 2005).

The Ottawa Charter for Health Promotion has made empowerment a key issue in the theory of health-promotion, which focuses on positive health enhancement rather than only ill-health prevention, mainly through improving the social conditions (Labonte, 1994). However, not many researchers have discussed the idea of empowerment in e-health system. Therefore this paper attempts to fill in the gaps which directly propose definition and scope of empowerment in healthcare organizations. In the next section we will discuss the concept of e-health services.

**E-HEALTH**

In light of Web technology within healthcare service, Eysenbach (2001) defined e-health as an emerging field in the intersection of medical informatics, public health and business, referring to health services and information delivered or enhanced through the Internet and related technologies. In a broader sense, the term characterizes not only a technical development, but also a state of mind, a way of thinking, an attitude, and a commitment for network, global thinking, to improve health care locally, regionally, and worldwide by using ICT.

Taking an advantage of the Internet in the area of health service which seems to be more reliable than traditional healthcare, the current status of e-health varies among countries. There are 193 countries that are a member of World Health Organization (WHO) in 2009; 114 of them participated in the global survey on e-health (WHO, 2011). Most developed countries have fully utilized e-health in their systems as they have the resources, expertise, and capital to implement them. While developing countries e-health systems have not been fully developed yet.

Several examples of countries that have implemented e-health are Canada, Singapore and Australia to name a few. Canada has established e-health Ontario on March 2009 with three targeted strategies to improve; diabetes management, medication management and wait times. One of the examples of the service offered is ePrescribing under the medication management. It authorizes and transmits prescriptions from physicians and other prescribers to pharmacists and other dispensers (e-Health Ontario, 2009). It prevents medication error due to illegible prescribing and reduces fraud prescriptions. "Participating prescribers and pharmacies at both sites will continue electronically prescribing until a provincial Medication Management System is in place" (e-Health Ontario, 2009).

In order to achieve the aims and benefits of e-health there must be a strategy in managing relationship between a healthcare provider and its customers as users. A good relationship between a healthcare provider and its customers will create a greater mutual understanding, trust, and patient involvement in decision making. Tool and technology in managing relationship with customers is well known as Customer Relationship Management (CRM). With the recent development in ICT, especially the emerging Web 2.0 technology such as social networks (e.g. Facebook, Twitter), blogs, wikis, and video sharing (Youtube), the capability of current CRM technology can be enhanced further. Greenberg (2009) proposes the new CRM with the Web 2.0 technology as Social CRM or CRM 2.0. The term of Social CRM and CRM 2.0 is used interchangeably. Both share new special capabilities of social media and social networks that provide powerful new approaches to surpass traditional CRM. Cipriani (2008) described the fundamental changes CRM 2.0 is introducing to the current, traditional CRM in terms of landscape. The most significant feature of CRM 2.0 is the network among customers and healthcare providers. This network creates value of network such as multi-ways communications and sharing of experience and knowledge. The study discusses the model that will highlight some pivotal characteristics of empowerment that need to be adopted in e-health business process through CRM 2.0.

**METHODOLOGY**

There are five stages to accomplish the study (see figure 1). These are literature review, reference model, survey, prototype, and testing. Research design is started from analyzing previous studies on the relating topics, and then we propose reference model as foundation for the next step. Further, questionnaires were derived from the reference model.

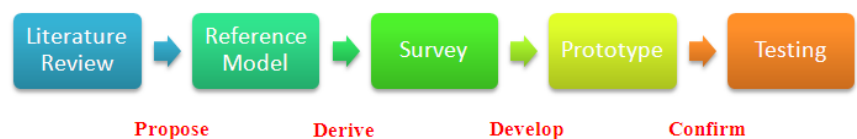

Fig. 1 Research Design

The aim of survey is to understand the customers' perspective on the features of proposed model, and the result of survey treated as preliminary requirements in developing prototype. At the end of whole research, we test the prototype in real healthcare scenario. However, the scope of this paper focuses on discussion of reference model.

**DISCUSSION**

A good relationship between a healthcare provider and its customers does not only improve customer's satisfaction, but also helps in fostering effective communications between them. Communications in healthcare industry is intense; we argue that our model will improve the relationship between healthcare providers and their customers for mutual benefit. The model operates in the area of healthcare organization–patient relationships inclusive with social network interaction, and how they can possibly share information to achieve health outcomes. Figure 2 is a proposed model of Social CRM in a healthcare environment. It offers a starting point for identifying possible theoretical mechanisms that might account for ways in which e-health business process can be understood.

      The framework is developed from Enterprise Social Networks, Internal Social Networks, Listening tool interfaces, Social CRM systems within healthcare provider, and healthcare value configuration (value chain and value shop). The aim to put internal and external social networks are to engage patients and export ideas, foster innovations of new services, quick response/feedback for existing service and technologies from people inside and outside organization (Anshari and Almunawar, 2012). Both provide a range of roles for patient or his/her family. The relationships can create emotional support, substantial aid and service, influence, advice and information that a person can use to deal with a problem. In addition, listening tool between Social Networks and CRM systems is a mechanism to capture actual data from social media and propagates this information forward to the CRM (Anshari et al., 2012). Social CRM empowers patient/family to have the ability in controlling his own data. Once patient/family registers to have service from healthcare provider, it will enable them to have personalized e-health systems with Social CRM as the frontline of the system. The system will authorize for each patient then; the authorization and self-managed account/service are granted to access all applications and data offered by the systems. This

authorization is expected to be in the long run since the information and contents continue to grow.

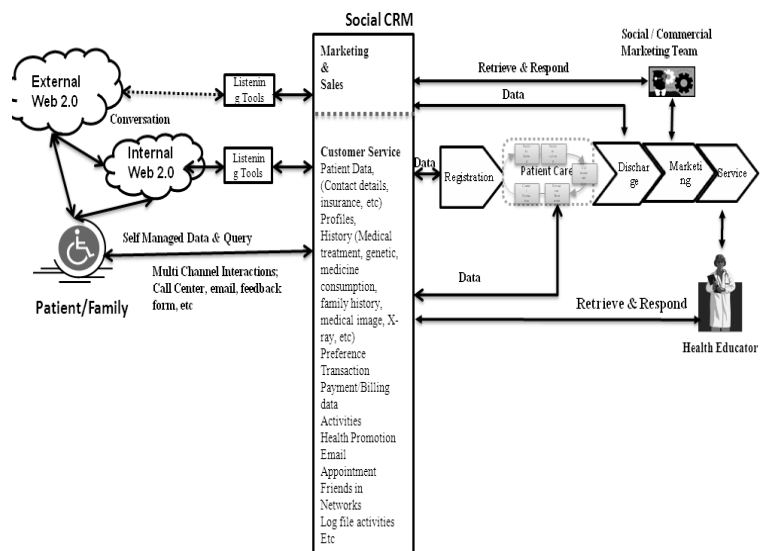

Fig. 2 Framework of Social CRM in Healthcare Organization

**E-Health's Business Process and Empowerment**

The framework in Figure 2 is extended into details healthcare's business process. The model is proposed to give an idea and comprehensive understanding on healthcare scenario. As mentioned in previous section that healthcare process implements the concept of value shop. The process of patient care is elaborated according to the value shop where value is created by mobilizing resources and activities to resolve a particular patient's problem. In other word, the healthcare business process is in circular form as shown in Figure 3. It explains interaction between personal and personal – provider either physically presence or indirectly service where patient may perform health related activities or interaction with healthcare provider through online. In short, e-health's business process comprise of interrelated online activities of personal daily life, clinical activities, and social activities that can be conducted anytime and anywhere.

Furthermore, figure 3 is referred as a reference model of Personal Health Cycle (PHC) with the modules of each stage. PHC comprises all activities of patient directly or indirectly relate to health service including extendable e-health service. PHC is composed from Personal Daily Life Activity (PDLA), Clinical Activities (CA); Checkups, In Patient Treatment, and Out Patient Treatment), and Social Life Activities (SLA). The PHC is derived from the framework of Social CRM in Healthcare in figure 2. Details of framework were discussed by Anshari and Almunawar (2011) which includes some important features of Social CRM such as customer's empowerment, social interactivity between healthcare organization–patients, and patients-patients. The framework offers new perspective in building relationships between healthcare organizations and customers and among customers in e-health scenario.

As shown from figure 3, each stage is composed from several modules. For instance, PDLA is consisted of Identity/Profile (ID), Personal Habits of patient (HB), Exercise activities (EX), Spiritual and Emotional activities (SE), Personal Health Plan (HP), Personal Account information (AC), and so on. Because of the approach is modular, each stage is extendible depends on the need of healthcare providers. Therefore from this modular approach, every healthcare provider varies in empowering their customers. Yellow circle line in module (figure 1) indicates that provider empower their customers to have control on that specific module, while dash circle line indicates that provider only give partial empowerment to customers like view EMR only. And no circle line means the healthcare providers provide no empowerment to the customers.

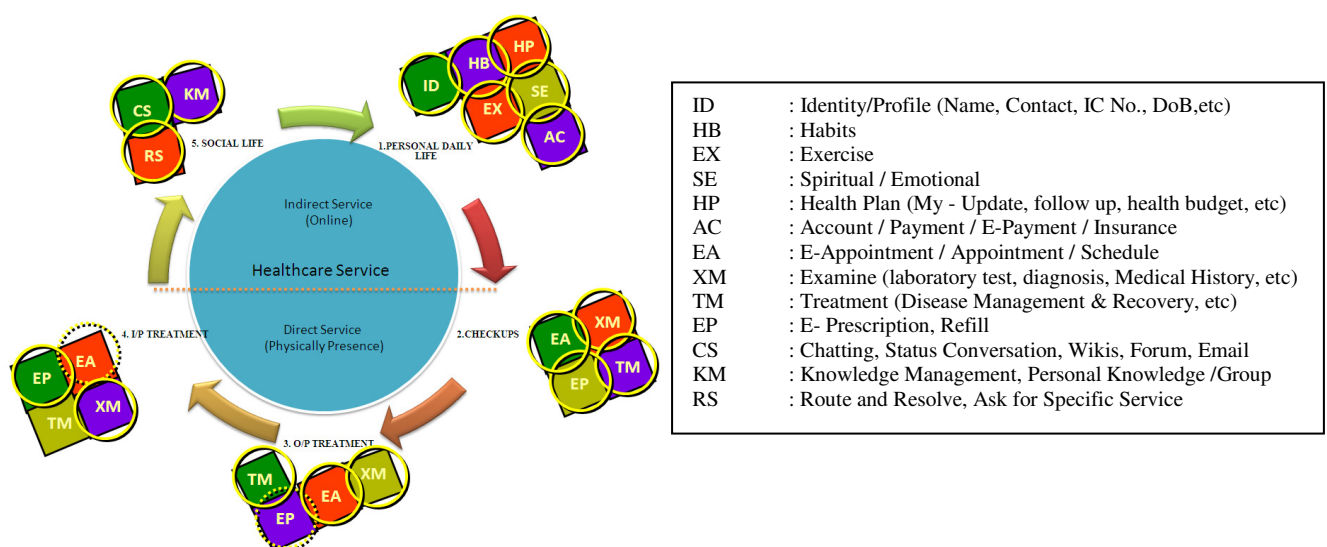

Fig. 3 Reference Model of Personal Health Cycle (PHC)

The model helps to define the empowerment and its scope in e-health system. Many researchers define empowerment in healthcare organizations partially. PLC is the central notion on how we define empowerment in healthcare organization. We analysis customers empowerment in healthcare organization including extended e-health service is an integrated healthcare process. Empowerment in e-health system is activity of customers personally, socially, and clinically which may have direct or indirect effect to their health status to have certain level of control towards their health related information depends on the policy of the organization so they can actively participate in the process of decision making to achieve their own health goals and others (Anshari et al., 2012). The PHC implies that the level of empowerment in any healthcare organization may vary depending upon the policy. For instance, healthcare organization may provide empowerment level to their customers only in PDLA level. Australia, on July 2012, will empower their citizen to the level of clinical activities (CA) with the authorization of viewing their own clinical information (NEHTA, 2012).

The reference model determines the dimension of healthcare business process with the possible perspective of empowerment. PHC can also provide a picture of the type of interaction in healthcare organizations. Since, PHC is in cycle, this means interaction is interconnected between patient-provider, patient-patient, and patient alone. With the Web 2.0 technology, all type of interaction can generate conversation and support that will benefit not only to customers but also to providers in creating strategy of sales, marketing, and customer service.

What is worthy to note from figure 3 is that to avoid ambiguity, we have defined and, in fact, differentiate Electronic Health Record (EHR) from Electronic Medical Record (EMR). EHR is a digital format and record of an individual which comprises all health related activities comprises of personal daily life, checkups, outpatient treatment, inpatient treatment, and social life with the purpose to give comprehensive personal health's views and history. In other word, EHR is comprises of PDLA, CA, and SLA. The EHR is not merely as an online tool to improve the efficiency and effectiveness of business process but also core components of personal quality to achieve continuity of health promotion, education, literature, knowledge management, and research development. Empowerment of EHR is the availability of personal health information to all possible parties including patients based on the role and urgency with the details of access control list to preserve confidentiality and privacy. EMR in e-health scenario is only part of the EHR as shown from figure 1. Additionally, checkups, inpatient treatment, and outpatient are the activities in CA's module that customers mostly have direct interaction with the provider for consultation or physical treatment (Anshari and Almunawar, 2012).

CONCLUSION

The empowerment can create a greater mutual understanding, trust, and patient involvement in decision making. Empowerment takes place in many ways and levels from the lowest level of digitalizing medical record to the level where customers can access to their Electronic Health Record (EHR). Empowerment can be in the form of interaction between customers to provider like accessing online service, booking online consultation, paying online, etc. Patient-patient interactions are also in need of empowerment which enables them to generate collaborative conversations in order to provide mutually beneficial value in trusted environment. In order to gain comprehensive views of customers' empowerment through e-health, we proposed and discussed the concept model of Personal Health Cycle (PHC). The PHC is an e-health business process that accommodates personal health-related activities, clinical activities, and social life activities as holistic approach to understand individual (patient) health status. The model is composed from various sub modules, and they can be extended depends on requirements and policies each healthcare provider. In conclusion, the model helps healthcare providers in determining which modules and sub modules that they are going to empower their customers when they implement e-health systems.